\documentclass[conference]{IEEEtran}
\IEEEoverridecommandlockouts
\usepackage{cite}
\usepackage{amsmath,amssymb,amsfonts}
\usepackage{algorithm,algorithmic}
\usepackage{graphicx}
\usepackage{textcomp}
\usepackage{xcolor}
\usepackage{bm}
\usepackage{lipsum}
\newcommand\blfootnote[1]{%
	\begingroup
	\renewcommand\thefootnote{}\footnote{#1}%
	\addtocounter{footnote}{-1}%
	\endgroup
}

\begin{document}

\title{Massive Unsourced Random Access for Massive MIMO Correlated Channels}

\author{\IEEEauthorblockN{Xinyu Xie, Yongpeng Wu, Junyuan Gao, and Wenjun Zhang}}

\maketitle

\begin{abstract}
This paper investigates the massive random access for a huge amount of user devices served by a base station (BS) equipped with a massive number of antennas. We consider a grant-free unsourced random access (U-RA) scheme where all users possess the same codebook and the BS aims at declaring a list of transmitted codewords and recovering the messages sent by active users. Most of the existing works concentrate on applying U-RA in the oversimplified independent and identically distributed (i.i.d.) channels. In this paper, we consider a fairly general joint-correlated MIMO channel model with line-of-sight components for the realistic outdoor wireless propagation environments. We conduct the activity detection for the emitted codewords by performing an improved coordinate descent approach with Bayesian learning automaton to solve a covariance-based maximum likelihood estimation problem. The proposed algorithm exhibits a faster convergence rate than traditional descent approaches. We further employ a coupled coding scheme to resolve the issue that the dimensions of the common codebook expand exponentially with user payload size in the practical massive machine-type communications scenario. Our simulations reveal that to achieve an error probability of $ 0.05 $ for reliable communications in correlated channels, one must pay a $ 0.9 $ to $ 1.3 $ dB penalty comparing to the minimum signal to noise ratio needed in i.i.d. channels on condition that a sufficient number of receiving antennas is equipped at the BS.
\end{abstract}


{\blfootnote{X. Xie, Y. Wu, J. Gao, and W. Zhang are with the Department of Electronic Engineering, Shanghai Jiao Tong University, Minhang 200240, China (e-mail: \{xinyuxie, yongpeng.wu, sunflower0515, zhangwenjun\}@sjtu.edu.cn) (Corresponding author: Yongpeng Wu).
		
The work of Y. Wu is supported in part by the National Key R\&D Program of China under Grant 2018YFB1801102, JiangXi Key R\&D Program under Grant 20181ACE50028, National Science Foundation (NSFC) under Grant 61701301, the open research project of State Key Laboratory of Integrated Services Networks (Xidian University) under Grant ISN20-03, and Shanghai Key Laboratory of Digital Media Processing and Transmission (STCSM18DZ2270700). The work of W. Zhang is supported by Shanghai Key Laboratory of Digital Media Processing and Transmission (STCSM18DZ2270700).}}

\vspace{-1em}
\section{Introduction}

The next generation of cellular technologies will proliferate in wirelessly connecting sensors, machines, and wearable biomedical devices, thereby form the architecture of the Internet of things (IoT). Massive machine-type communications (mMTC) is a representative IoT application scenario where a base station (BS) equipped with multiple antennas renders service to a large number of machine-type users \cite{101}. Typically, in a certain period of time, only a small fraction of users within the system are operational and transmit abbreviated messages in the order of $ 100 $ bits. The characters of sporadic traffic patterns and small data payloads raise higher demands for more efficient multiple-access schemes that aim to accommodate the burgeoning number of users.

Employing conventional grant-based random access (RA) schemes to mMTC systems would manifest high energy consumption and latency, thus, grant-free RA protocols have recently attracted significant attention, where users send data immediately to the BS without approvals. A typical type of grant-free RA scheme is based on the allocation of pilot signals \cite{102,1021,103}, where unique pilots are randomly selected by active users for activity detection (AD) and channel estimation (CE) before data transmission. However, the confined pilot resources limit the support for more potential users.

As a prospective alternative, a novel modality of grant-free \emph{unsourced random access} (U-RA) is introduced \cite{104,105}, where users compulsively utilize the same codebook and the BS only needs to acquire a list of transmitted messages without the obligation of associating them to the specific active users. Although it is found in \cite{104} that existing approaches like ALOHA and CDMA perform poorly under U-RA, lots of revised coding schemes are proposed to promote the properties \cite{106,107,108}. Recently, the application of U-RA is ulteriorly extended to the block-fading massive MIMO channel \cite{109,110}, where modified coupled coding schemes similar to \cite{108} is proposed to overcome the problem that the dimensions of the coding matrix grow exponentially with the user payloads. The inner decoder at the receiver side in \cite{109} conducts AD to the transmitted codewords following the multiple-measurement vector (MMV) CS recovery paradigm and employs the \emph{approximate message passing} (AMP) algorithm \cite{102}. Nevertheless, a covariance-based maximum likelihood (ML) estimation scheme is adopted in \cite{110} to instantiate the AD problem, and the corresponding coordinate descent (CD) solution \cite{111} surpasses the performance of AMP. Also, the AMP algorithm must regard the large-scale fading coefficients (LSFCs) as either deterministic known parameters or random parameters with known prior distribution information. In contrast, the covariance-based scheme and its CD solution treat the LSFCs unknown to the system and avoid the trouble of individually measuring these parameters. However, the aforementioned literature only considers independent and identically distributed (i.i.d.) massive MIMO channels, which may be impractical because, in realistic outdoor wireless propagation environments, antennas at both the BS and user terminal are subjected to high correlations. Furthermore, the CD method in \cite{111} practices a random coordinate selection policy and reveals a slow convergence rate during actual operations.

Concentrating on the U-RA scheme, the main objective of this paper is to realize reliable communications between a large number of machine-type users and the massive MIMO BS under spatially correlated channels. Toward this end, we consider a rather generic joint-correlation channel model that takes line-of-sight (LOS) components into account. Only a small fraction of all potential users are active and transmit corresponding codewords from a common codebook using coherent blocks. We follow the paradigm of the covariance-based recovery problem for AD of the emitted codewords, where the BS and users only have access to partial information about the correlated channels and treat arguments like LSFCs unknown. We introduce an improved coordinate-wise descent algorithm with \emph{Bayesian learning automaton} (BLA) for ML estimation to achieve a faster convergence rate than conventional CD approaches in \cite{110,111}. We further resort to a concatenated coding strategy to reduce the size of the common codebook. Exhaustive simulations indicate that to satisfy the error probability needed for feasible communications (e.g., error probability $ P_{\text{e}}<0.05 $), there is a $ 0.9 $ to $ 1.3 $ dB gap between the SNR required in our proposed scheme under correlated channels and that for i.i.d. channels when different numbers of antennas are set at the BS.

Throughout this paper, scalar variants and constants are denoted by non-boldface letters, while bold lowercase and uppercase letters denote column vectors and matrices, respectively. $ [\mathbf{Y}]_{mn} $ indicates the $ (m,n) $-th entry of the matrix $ \mathbf{Y} \in \mathbb{C}^{M \times N} $ and $ \mathbf{y}_n $ denotes the $ n $-th column of $ \mathbf{Y} $. We signify the conjugate, transpose, and conjugate transpose by superscripts $ (\cdot)^*,(\cdot)^T $, and $ (\cdot)^H $, respectively. The operator $ \odot $ denotes the Hadamard product of two matrices with the same size. We denote the Euclid norm of vector $ \mathbf{y} $ and the Frobenius norm of matrix $ \mathbf{Y} $ by $ \left\| \mathbf{y} \right\|_2  $ and $ \left\| \mathbf{Y} \right\|_{\text{F}} $, respectively. The mathematical expectation operator is represented by $ \mathbb{E} \{\cdot\} $. $ |\mathcal{X}| $ calculates the number of elements in set $ \mathcal{X} $ and $ \mathcal{X} \backslash \mathcal{Y} $ represents the set $ \{z:z \in \mathcal{Y},z \notin \mathcal{X}, \mathcal{X} \subseteq \mathcal{Y} \} $. $ [M] $ represents the set of integers $ \{1,2,\dots,M\} $.

\section{Channel and System Model}

\subsection{Correlated Channel Model}

We consider a single-cell massive MIMO network system comprised of a total number of $ K_{\text{tot}} $ multiple-antenna users. They communicate to a single BS equipped with a massive number of $ M $ antennas through the uplink synchronizing scheme. We denote the set of all users within the system by $ \mathcal{K}_{\text{tot}} $ with $ |\mathcal{K}_{\text{tot}}|=K_{\text{tot}} $. Each user $ k \in \mathcal{K}_{\text{tot}} $ is provided with $ N_k $ transmitting antennas and $ N=\sum_{k=1}^{K_{\text{tot}}} N_k $ counts the sum amount of transmitting antennas at the user terminal.

We consider a rather general joint-correlated MIMO channel between user $ k \in \mathcal{K}_{\text{tot}} $ and the BS \cite{201}, the $ M \times N_k $ channel matrix $ \mathbf{H}_k $ is modeled as
\begin{align}
	\mathbf{H}_k &= \mathbf{U}_{\text{r},k} \tilde{\mathbf{H}}_k \mathbf{U}_{\text{t},k}^{H} \notag \\ 
	&=\mathbf{U}_{\text{r},k} \left( \bar{\mathbf{H}}_k+\mathbf{P}_k \odot \hat{\mathbf{H}}_k \right) \mathbf{U}_{\text{t},k}^{H}, \label{cochannel}
\end{align}
where $ \tilde{\mathbf{H}}_k = \bar{\mathbf{H}}_k+\mathbf{P}_k \odot \hat{\mathbf{H}}_k $, $ \mathbf{U}_{\text{r},k} \in \mathbb{C}^{M \times M} $ and $ \mathbf{U}_{\text{t},k} \in \mathbb{C}^{N_k \times N_k} $ are deterministic unitary matrices, $ \bar{\mathbf{H}}_k \in \mathbb{C}^{M \times N_k} $ is a deterministic matrix modeling the LOS components with at most one nonzero element in each row and each column, $ \mathbf{P}_k \in \mathbb{R}^{M \times N_k} $ is a deterministic matrix with real-valued nonnegative entries, and $ \hat{\mathbf{H}}_k \in \mathbb{C}^{M \times N_k} $ is a random matrix whose entries are i.i.d complex variables with zero-mean. We normalize $ \mathbf{H}_k $ such that
\begin{align}
	\mathbb{E} \left\lbrace \text{tr} \left( \mathbf{H}_k\mathbf{H}_k^H \right) \right\rbrace = N_kM. \label{Hnorm}
\end{align}

For convenience but without loss of generality, we assume that nonzero elements in $ \bar{\mathbf{H}}_k $ have indices of $ \left( d,d \right)  $ for $ 1 \leq d \leq \min \left( N_k,M \right)  $. We further define
\begin{align}
	\mathbf{\Omega}_k = \mathbb{E} \left\lbrace \tilde{\mathbf{H}}_k \odot \tilde{\mathbf{H}}_k^* \right\rbrace = \bar{\mathbf{H}}_k \odot \bar{\mathbf{H}}_k + \mathbf{P}_k \odot \mathbf{P}_k. \label{Omega}
\end{align}
The $ (m,n) $-th entry of $ \mathbf{\Omega} $, i.e., $ [\mathbf{\Omega}_k]_{mn} $, describes the average power coupling between the $ m $-th receive eigenmode and $ n $-th transmit eigenmode. To be consistent with the normalization of $ \mathbf{H}_k $ in \eqref{Hnorm}, the power constraint of \eqref{Omega} can be expressed as
\begin{align}
	\sum_{m=1}^{M}\sum_{n=1}^{N_k} \left[ \mathbf{\Omega}_k \right]_{mn} = N_kM.
\end{align}

The transmit and receive correlation matrices of $ \mathbf{H}_k $ are
\begin{align}
	\mathbf{R_{\text{t}}} &= \mathbb{E} \left\lbrace \mathbf{H}_k^H\mathbf{H}_k \right\rbrace = \mathbf{U}_{\text{t},k}\mathbf{\Lambda}_{\text{t},k}\mathbf{U}_{\text{t},k}^H, \label{trans corr}\\
	\mathbf{R_{\text{r}}} &= \mathbb{E} \left\lbrace \mathbf{H}_k\mathbf{H}_k^H \right\rbrace = \mathbf{U}_{\text{r},k}\mathbf{\Lambda}_{\text{r},k}\mathbf{U}_{\text{r},k}^H,
\end{align}
where $ \mathbf{\Lambda}_{\text{t},k} $ and $ \mathbf{\Lambda}_{\text{r},k} $ are diagonal matrices with $ \left[ \mathbf{\Lambda}_{\text{t},k} \right]_{nn} = \sum_{m=1}^{M} \left[ \mathbf{\Omega}_k \right]_{mn}$ for $ n \in [N]$ and $ \left[ \mathbf{\Lambda}_{\text{r},k} \right]_{mm} = \sum_{n=1}^{N_k} \left[ \mathbf{\Omega}_k \right]_{mn} $ for $ m \in [M]$, respectively. In massive MIMO environments, these covariance matrices tend to be low-rank with a small number of significant eigenvalues \cite{202}, indicating that many entries in $ \bar{\mathbf{H}}_k $ and $ \mathbf{P}_k $ are approximately zero. If further postulate that the scatters between different users and the BS are independent with each other, we have
\begin{equation}
	\lim\limits_{M,N_k,N_{k'} \to \infty} \mathbb{E} \left\lbrace \tilde{\mathbf{H}}_k^H\tilde{\mathbf{H}}_{k'} \right\rbrace = \mathbf{0}_{N_k \times N_{k'}}. \label{Horth}
\end{equation}

If $ M $ is sufficiently large, the eigenvectors in $ \mathbf{U}_{\text{r},k} $ for different users tend to be equal, i.e., $ \mathbf{U}_{\text{r},k}=\mathbf{U}_{\text{r}} $ for $ k \in \mathcal{K}_{\text{tot}} $. In this paper, we stick to the assumption that each user only has the knowledge of $ \mathbf{U}_{\text{t},k} $ and the BS only knows the statistical parameter $ \mathbf{U}_{\text{r}} $. Such information between a transmitter-receiver pair can be obtained over time utilizing averaged uplink and downlink channel measurements in Time Division Duplexing (TDD) systems using channel reciprocity \cite{203}. Since the information is much more robust to minor changes in the surroundings and remains valid for a longer time than the channel state information (CSI), we only need to update it at relatively long intervals.

Note that the channel model described in \eqref{cochannel} encompasses many of the existing statistical MIMO channel models. For instance, if $ \bar{\mathbf{H}}_k = \mathbf{0} $ and $ \mathbf{P}_k $ is a rank-one matrix, the Kronecker model is retrieved \cite{204,205}. Setting $ \mathbf{P}_k $ to have arbitrary rank while fixing $ \mathbf{U}_{\text{t},k} $ and $ \mathbf{U}_{\text{r},k} $ to be Fourier matrices, the virtual channel representation for uniform linear arrays (ULA) is recovered \cite{206}. If further let $ \mathbf{U}_{\text{t},k} $ and $ \mathbf{U}_{\text{r},k} $ to be arbitrary unitary matrices, we acquire the Weichselberger’s channel model \cite{207}.

\subsection{System Model}

We consider a block-fading channel with coherent blocks of $ D $ signal dimensions where the channel parameters remain constant. There are $ K_{\text{tot}} $ users within the system but only a small set of $ K_{\text{a}} $ users denoted by $ \mathcal{K}_{\text{a}} $ transmit messages synchronously. We assume that each active user $ k \in \mathcal{K}_{\text{a}} $ is appointed to send a $ W $-bit message $ \mathbf{m}(k)=\mathbf{m}^T_{i_k} $ from the same message set $ \mathcal{M}=\{\mathbf{m}_{i_k}:i_k \in [2^{W}]\} $. Each message $ \mathbf{m}_{i_k} $ corresponds to a codeword to be emitted over $ S $ entire coherent blocks during $ S $ successive time slots, such that each codeword has the length of $ C = DS $. We abide the framework of U-RA in \cite{104} where all users own a $ C \times 2^{CR} $-sized common codebook with $ R = W/C $ being the coding rate. The objective of the BS is to produce a list $ \mathcal{L}=\left\lbrace \mathbf{m}(k):k \in \mathcal{K}_{\text{a}} \right\rbrace $ of transmitted messages during these blocks without matching them to the original users. In the U-RA scheme, the error event probability is defined on the per-user basis in contrast to global for all users, i.e., \textit{per-User probability of misdetection} and \textit{probability of false-alarm}, expressed respectively as
\begin{align}
p_{\text{md}} &= \dfrac{1}{K_{\text{a}}} \sum_{k \in \mathcal{K}_{\text{a}}} \mathbb{P} \left( \mathbf{m}(k) \notin \mathcal{L} \right), \label{pmd} \\
p_{\text{fa}} &= \dfrac{\left| \mathcal{L} \backslash \left\lbrace \mathbf{m}(k):k \in \mathcal{K}_{\text{a}} \right\rbrace \right| }{\left| \mathcal{L} \right| }. \label{pfa}
\end{align}	

We provisionally assume that the codewords corresponding to the messages generated from active users are transmitted using only one block, i.e., $ S=1 $ and $ C=D $. All potential users are allotted to the same coding matrix $ \mathbf{A} \in \mathbb{C}^{D \times 2^{CR}} = [\mathbf{a}_1,\dots,\mathbf{a}_{2^{CR}}]  $ with each column $ \mathbf{a}_i, i \in [2^{CR}] $ representing a codeword whose entries are taken from the i.i.d. complex Gaussian distribution with zero mean and unit variance, such that $ \left\| \mathbf{a}_i \right\|_2^2 = D $. If a specific active user $ k \in \mathcal{K}_{\text{a}} $ wishes to send a message $ \mathbf{m}(k)=\mathbf{m}_{i_k} $, the corresponding codeword to be emitted is drawn from the $ i_k $-th column of $ \mathbf{A} $. Note that the unitary matrix $ \mathbf{U}_{\text{t},k} $ is posited to be known to the users, hence we directly employ it as the precoding matrix. The sending signals at all transmitting antennas can be written as
\begin{align}
	\mathbf{X}_k &= \mathbf{AB}_k\mathbf{U}_{\text{t},k}^T,
\end{align}
where $ \mathbf{B}_k \in \left\lbrace 0,1 \right\rbrace ^{2^{CR} \times N_k} $ is a binary matrix with each row being all-zero but ones in the $ i_k $-th row for $ k \in \mathcal{K}_{\text{a}} $, while contains all zeros for $ k \in \mathcal{K}_{\text{tot}} \backslash \mathcal{K}_{\text{a}} $.

Having the knowledge of $ \mathbf{U}_{\text{r}} $, the received signals from all users at the BS with $ M $ receiving antennas can be modeled as
\begin{align}
	\mathbf{Y} &= \left[ \mathbf{X}_1\mathbf{G}^{\frac{1}{2}}_1 \cdots \mathbf{X}_{K_\text{tot}}\mathbf{G}^{\frac{1}{2}}_{K_\text{tot}} \right] 
	\left[ \begin{matrix}
	\mathbf{H}^T_1 \\
	\vdots \\
	\mathbf{H}^T_{K_\text{tot}}
	\end{matrix} \right]\mathbf{U}_{\text{r}}^* + \mathbf{Z}\mathbf{U}_{\text{r}}^* \notag \\
	&= \mathbf{ABG}^{\frac{1}{2}}\tilde{\mathbf{H}} + \tilde{\mathbf{Z}}, \label{sys model}
\end{align}
where $ \mathbf{G}_k=g_k\mathbf{I}_{N_k} $ for $ k \in \mathcal{K}_{\text{tot}} $ is the matrix of (unknown) LSFCs with $ g_k $ measuring the average antenna transmission power, $ \mathbf{G}=\text{diag}\left[ \mathbf{G}_1,\dots,\mathbf{G}_{K_\text{tot}} \right] $ is a diagonal matrix of order $ N $, $ \mathbf{B} = \left[ \mathbf{B}_1,\dots,\mathbf{B}_{K_\text{tot}} \right] \in \{0,1\}^{2^{CR} \times N} $, $ \tilde{\mathbf{H}}=[\tilde{\mathbf{H}}_1,\dots,\tilde{\mathbf{H}}_{K_\text{tot}}]^T \in \mathbb{C}^{N \times M} $, $ \mathbf{Z} \in \mathbb{C}^{D \times M} $ is the matrix of additive white Gaussian noise with entries generated from the i.i.d. complex Gaussian distribution, i.e., $ \left[ \mathbf{Z} \right]_{dm} \sim \mathcal{CN} \left(0,\sigma^2 \right) $, and $ \tilde{\mathbf{Z}}=\mathbf{Z}\mathbf{U}_{\text{r}}^* $ is the matrix of the equivalent noise samples. Note that we assume the LSFCs of all antennas of a certain user are the same since the signals transmitted by these antennas take similar paths to the BS on the macro scale.

\section{Covariance Based Activity Detection}

\subsection{Problem Formulation}
Based on the independency of coupling matrices $ \tilde{\mathbf{H}}_k, k \in \mathcal{K}_\text{tot} $ between different users \eqref{Horth}, the covariance matrix of $ \tilde{\mathbf{H}} $ in \eqref{sys model} can be approximated by
\begin{align}
\mathbb{E}\left\lbrace \tilde{\mathbf{H}}\tilde{\mathbf{H}}^H \right\rbrace = \text{diag}\left[ \mathbf{\Lambda}_{\text{t},1},\dots,\mathbf{\Lambda}_{\text{t},K_\text{tot}} \right] = \mathbf{\Lambda}, \label{tildeHorth}
\end{align}
where $ \mathbf{\Lambda} \in \mathbb{C}^{N \times N} $ is a block diagonal matrix with each block $ \mathbf{\Lambda}_{\text{t},k} \in \mathbb{C}^{N_k \times N_k}$ having the expression as \eqref{trans corr}.

We make a crucial observation from \eqref{sys model} and \eqref{tildeHorth} that the columns $ \mathbf{y}_m $ of $ \mathbf{Y}, m \in [M] $, can be equivalently viewed as independent samples from a multivariate complex Gaussian distribution, i.e., $ \mathbf{y}_m \sim \mathcal{CN}\left( 0,\mathbf{\Sigma_y} \right) $. Note that similar observation is made under i.i.d. channels based on the spatially white user channel vectors \cite{110}. By computing $ \mathbb{E}\left\lbrace \mathbf{y}_m\mathbf{y}_m^H \right\rbrace  $, the covariance matrix $ \mathbf{\Sigma_y} $ takes on the form
\begin{align}
	\mathbf{\Sigma_y} &= \mathbf{ABG}^{\frac{1}{2}}\mathbf{\Lambda}( \mathbf{G}^{\frac{1}{2}})^H\mathbf{B}^H\mathbf{A}^H \notag \\
	&= \mathbf{A\Gamma A}^H + \sigma^2\mathbf{I}_D, \label{cory}
\end{align}
where $ \mathbf{\Gamma} = \mathbf{BG}^{\frac{1}{2}}\mathbf{\Lambda}(\mathbf{G}^{\frac{1}{2}})^H\mathbf{B}^H = \text{diag}\left[ \gamma_1,\dots,\gamma_{2^{CR}} \right] $ with $ \gamma_i=\sum_{n=1}^{N}[\mathbf{B}]_{in}^2[\mathbf{G}]_{nn}[\mathbf{\Lambda}]_{nn} $, $ i \in [2^{CR}] $. Notice that $ \gamma_i $ takes nonzero value only when the $ i $-th row of $ \mathbf{B} $ has nonzero elements, in other words, at least one active user transmits the $ i $-th codeword. Hence, we define $ \bm{\gamma} = \left[ \gamma_1,\dots,\gamma_{2^{CR}} \right] $ as the activity pattern of the codewords. Note that since the elements of $ \mathbf{G} $ are embedded in $ \bm{\gamma} $ to be estimated, we do not need any information about the LSFCs.

In order to determine which codewords are transmitted during the time slot, the BS is supposed to recover $ \bm{\gamma} $ from the observed sample covariance $ \widehat{\mathbf{\Sigma}}_{\mathbf{y}} = \frac{1}{M}\mathbf{YY}^H $ of the received signals. It is hardly possible for the estimator at the BS to gain knowledge of the active user number $ K_{\text{a}} $ due to the sporadic traffic pattern in mMTC. Therefore, we adopt the hard decision strategy by pre-assigning suitable fixed thresholds at each time slot. The transmitted message list is recovered as
\begin{equation}
\mathcal{L} = \left\lbrace \mathbf{m}_i: \widehat{\bm{\gamma}}_i > \zeta, i \in [2^{CR}] \right\rbrace,
\end{equation}
where $ \zeta $ is the pre-assigned threshold and $ \widehat{\bm{\gamma}} = \left[ \widehat{\gamma}_1,\dots,\widehat{\gamma}_{2^{CR}} \right] $ is the estimation of $ \bm{\gamma} $.

\subsection{Maximum Likelihood Estimation}

The likelihood of $ \mathbf{Y} $ given $ \bm{\gamma} $ is expressed as
\begin{align}
	\mathbb{P}\left( \mathbf{Y}|\bm{\gamma} \right) &= \prod_{m=1}^{M}\dfrac{1}{\text{det}\left( \pi \mathbf{\Sigma_y} \right) } \exp\left( -\mathbf{y}_m^H\mathbf{\Sigma}^{-1}_\mathbf{y}\mathbf{y}_m \right) \notag \\
	&\overset{(a)}{=} \dfrac{1}{\left[ \text{det}(\pi \mathbf{\Sigma_y}) \right]^M } \exp\left[ -M\text{tr}\left( \mathbf{\Sigma_y}^{-1}\widehat{\mathbf{\Sigma}}_{\mathbf{y}} \right)\right] \label{likelihood},
\end{align}
where $ \text{det}(\cdot) $ and $ \text{tr}(\cdot) $ are the determinant and trace operations of a matrix, respectively, and $ (a) $ is derived based on the observation that the columns of Y are i.i.d.. By harnessing the log-likelihood cost function $ f\left( \bm{\gamma} \right) = -\frac{1}{M}\log \mathbb{P}\left( \mathbf{Y}|\bm{\gamma} \right) $, maximizing the likelihood in \eqref{likelihood} can be converted into resolving the following problem
\begin{subequations}\begin{align}
	\mathop{\text{minimize}}\limits_{\bm{\gamma}} \quad &f\left( \bm{\gamma} \right) = \log \text{det} \left( \mathbf{\Sigma_y} \right)  + \text{tr}\left( \mathbf{\Sigma}^{-1}_{\mathbf{y}}\widehat{\mathbf{\Sigma}}_{\mathbf{y}} \right) \label{MLa} \\
	\text{subject to} \quad & \bm{\gamma}\ge 0 \label{MLb}
\end{align}\end{subequations}

Note that in \eqref{likelihood}, $ p\left( \mathbf{Y}|\bm{\gamma} \right) $ depends on $ \mathbf{Y} $ only through the sample covariance matrix $ \widehat{\mathbf{\Sigma}}_{\mathbf{y}} $, therefore $ \widehat{\mathbf{\Sigma}}_{\mathbf{y}} $ is a sufficient statistic for the estimation of $ \bm{\gamma} $. For the same reason, we refer to our paradigm of estimation as the covariance-based approach. The constraint that $ \bm{\gamma}\ge 0 $ in \eqref{MLb} ensures the positive definiteness of the covariance matrix $ \mathbf{\Sigma_y} $. It is worth mentioning that leveraging $ \widehat{\mathbf{\Sigma}}_{\mathbf{y}} \in \mathbb{C}^{D \times D} $ instead of exploiting the row sparsity of $ \mathbf{Y} \in \mathbb{C}^{D \times M} $ by some compressed sensing approaches \cite{109} results in a remarkable dimensional reduction, making this approach peculiarly attractive for some massive MIMO systems where $ M>D $.

\subsection{Iterative Descent Algorithm}

Existing descent algorithms are well adapted to the problem of optimizing $ f(\bm{\gamma}) $ over a natural parameter space of $ \mathbb{R}^{2^{CR}}_+ $ with $ 2^{CR} $ dimensions. Paper \cite{111} suggests a CD algorithm where at each step $ f(\bm{\gamma}) $ is optimized in regard to only one of its objects $ \gamma_i, i \in [2^{CR}] $ abiding by the updating rules summarized in Algorithm 1, Line 16-18. After adequate iterations over the whole collection of coordinates, the optimization function will ultimately converge to its minimum. However, such an algorithm practices a random selection policy at the whole coordinate set, which may result in a low convergence rate. An intuitive explanation for this situation is that only a small section of arguments in $ \bm{\gamma} $ indicating the active codewords require plentiful rounds of estimation such that $ f(\bm{\gamma}) $ will precisely approach its minimum, while others can be vaguely estimated with a few rounds since they take small values and have limited effects on the overall results.

\begin{algorithm}[t]
	\caption{Activity Detection via Coordinate Descent}
	\begin{algorithmic}[1]
		\STATE \textbf{Input:} Sample covariance matrix $ \widehat{\mathbf{\Sigma}}_{\mathbf{y}}=\frac{1}{M}\mathbf{YY}^H $ of the received signal $ \mathbf{Y} $, interger $ Q $, $ Q_{\text{mod}} $.
		\STATE \textbf{Initialize:} $ \mathbf{\Sigma}=\sigma^2 \mathbf{I}_D $, $ \bm{\gamma}^0=\mathbf{0} $, and $ \alpha_1=\alpha_2=\beta_1=\beta_2=1 $.
		\FOR{$ q=1,2,\dots,Q $}
		\IF{$ q \mod Q_{\text{mod}}==1 $}
		\STATE Update $ \psi^q_i=r^q_i $ for $ i \in \left[ 2^{CR} \right] $.
		\ENDIF
		\STATE Generate $ \epsilon^q_1 \sim \text{Beta} (\alpha_1,\beta_1) $ and $ \epsilon^q_2 \sim \text{Beta} (\alpha_2,\beta_2) $.
		\STATE Select $ l_q = \mathop{\max}\limits_{l \in \{1,2\}} \left\lbrace \epsilon^{q}_1,\epsilon^{q}_2 \right\rbrace $.
		\STATE Generate $ z \sim \text{Bernoulli} (\epsilon^{q}_{l_q}) $.
		\STATE Update $ (\alpha_{l_q},\beta_{l_q})=(\alpha_{l_q}+z,\beta_{l_q}+1-z) $.
		\IF{$ z==1 $}
		\STATE Select $ i_q = \mathop{\max}\limits_{t \in [2^{CR}]} \left\lbrace \psi^q_1,\dots,\psi^q_{2^{CR}} \right\rbrace $.
		\ELSE
		\STATE Select $ i_q \in \left[ 2^{CR} \right] $ randomly.
		\ENDIF
		\STATE Set $ d=\max \left\lbrace \frac{\mathbf{a}_{i_q}^H\mathbf{\Sigma}^{-1}\widehat{\mathbf{\Sigma}}_{\mathbf{y}}\mathbf{\Sigma}^{-1}\mathbf{a}_{i_q}-\mathbf{a}_{i_q}^H\mathbf{\Sigma}^{-1}\mathbf{a}_{i_q}}{(\mathbf{a}_{i_q}^H\mathbf{\Sigma}^{-1}\mathbf{a}_{i_q})^2 },-\gamma_{i_q} \right\rbrace $.
		\STATE Update $ \gamma^{q}_{i_q}=\gamma^{q-1}_{i_q}+d $ and $ \gamma^{q}_{i}=\gamma^{q-1}_{i} $ for $ i \ne i_q $.
		\STATE Update $ \mathbf{\Sigma}^{-1}=\mathbf{\Sigma}^{-1}-\frac{d\mathbf{\Sigma}^{-1}\mathbf{a}_{i_q}\mathbf{a}_{i_q}^H\mathbf{\Sigma}^{-1}}{1+d\mathbf{a}_{i_q}^H\mathbf{\Sigma}^{-1}\mathbf{a}_{i_q}} $.
		\STATE Calculate $ r^{q+1}_{i_q} $ according to \eqref{reward}.
		\STATE Update $ \psi^{q+1}_{i_q}=r^{q+1}_{i_q} $ and $ \psi^{q+1}_i=\psi^{q}_i $ for $ i \ne i_q $.
		\ENDFOR 
		\STATE \textbf{Output:} The estimation $ \widehat{\bm{\gamma}} = \bm{\gamma}^{Q} $.
	\end{algorithmic} 
\end{algorithm}

In order to enhance the estimation accuracy after limited iterations, we introduce an alternative coordinate selection strategy illustrated as a multi-armed bandit (MAB) problem \cite{301}. At each iteration of descent, an arm representing a certain coordinate is pulled and produces a random reward. The reward function is derived by measuring the descent degree of $ f(\bm{\gamma}) $ when updating the coordinate $ i \in [2^{CR}] $ at the $ q $-th iteration and takes on the form
\begin{equation}
	r^q_i = \dfrac{d\mathbf{\Sigma}^{-1}\mathbf{a}_{i}\mathbf{a}_{i}^H\mathbf{\Sigma}^{-1}}{1+d\mathbf{a}_{i}^H\mathbf{\Sigma}^{-1}\mathbf{a}_{i}} - \log \left( 1+d\mathbf{a}_{i}^H\mathbf{\Sigma}^{-1}\mathbf{a}_{i} \right). \label{reward}
\end{equation}
The objective of the MAB problem is to maximize the accumulative sum of rewards received during $ Q $ iterations, i.e., $ \sum_{q=1}^{Q} r^q_{i_q} $ with $ i_q \in [2^{CR}] $ the chosen arm at the $ q $-th iteration. As information of the arms’ payouts are gathered, we are forced to choose between exploiting the arm that currently yields the largest reward and exploring other arms that may lead to higher rewards in the future. This dithering is resolved by the so-called $ \epsilon $-greedy exploration approach, where we take the probability $ \epsilon $ to perform the previous greedy action (see Algorithm 1, Line 12) and $ 1-\epsilon $ to the later completely random action (see Algorithm 1, Line 14). The appropriate setting of the Bernoulli distribution probability parameter $ \epsilon $ will lead to a tradeoff between \emph{exploration} and \emph{exploitation}.

Instead of simply assigning a fixed probability in advance, we suggest that dynamically optimizing the argument $ \epsilon $ can be formulated as a two-armed Bernoulli bandit (TABB) problem. In the TABB problem, one must ceaselessly extract one of the two arms and implement a Bernoulli trial based on the inherent information of the chosen arm, with each trial leading to a payout of either a reward $ 1 $ or a penalty $ 0 $. To maximize the unknown expected total payouts, we must achieve a tradeoff between exploiting existing knowledge about the arms and exploring new information. Many efficient methods have arisen to solve the problem and we are inclined to apply the BLA approach \cite{302}. By taking advantage of the potential Bayesian prior information of the arms, BLA exhibits a characteristic of self-correcting and converges to the best decision of choosing the optimal arm.

Following the process of BLA, two different Bernoulli distribution probability parameter $ \epsilon_1 $ and $ \epsilon_2 $ act as two arms in the TABB problem with each arm having the Bayesian prior satisfying the probability density function of a Beta distribution, i.e., $ \epsilon_l \sim \text{Beta}(\alpha_l,\beta_l) $ with $ f(x;\alpha_l,\beta_l) = \frac{\Gamma(\alpha_l+\beta_l)}{\Gamma(\alpha_l)+\Gamma(\beta_l)}x^{\alpha_l-1}(1-x)^{\beta_l-1} $ for $ l \in \{1,2\} $, where $ \Gamma(\cdot) $ is the Gamma function and $ \alpha_l,\beta_l>0 $. We initialize with $ \alpha_1=\alpha_2=\beta_l=\beta_2=1 $. At the $ q $-th iteration round, these two arms both generate random samples denoted by $ \epsilon^q_1$ and $\epsilon^q_2 $. We select the arm $ l_q \in \{1,2\} $ with the larger value and practice a Bernoulli trial with an outcome $ z \sim \text{Bernoulli}(\epsilon^q_{l_q}) $. Then the chosen arm renewals its distribution by updating parameters $ (\alpha^q_{l_q},\beta^q_{l_q}) $ to $ (\alpha^{q+1}_{l_q},\beta^{q+1}_{l_q})=(\alpha^q_{l_q}+z,\beta^q_{l_q}+1-z) $, resulting in a new arm chosen in the next round. As more observations are obtained, BLA will gradually prefer to select the arm which is most likely the optimal one.

We establish the framework of the CD algorithm with BLA in Algorithm 1. Note that we only renewal the whole set of $ \psi^q_i $, $ i \in [2^{CR}] $ each $ Q_{\text{mod}} $ rounds to reduce computation complexity, whereas in other rounds, $ \psi^q_i $ takes the most recently available value. We can also treat the noise variance $ \sigma^2 $ unknown and estimate it along with $ \bm{\gamma} $. In section V, we will perform a numerical experiment to demonstrate that our proposed algorithm achieves a higher convergence rate than the original CD approach in \cite{111}.

\section{Complexity Reduction via Coupled Coding}

In realistic communication systems, it is impractical to transmit messages within just one coherence block. Especially in U-RA, on account of the scale of the common codebook increasing proportionally to the message length, the system becomes impracticable even for relatively small-sized messages. To reduce the codebook size, we turn to the coupled coding scheme in \cite{108,110} which consists of inner and outer codes.

At the outer tree encoding step, the $ W $-bit message to be transmitted over $ S $ coherence blocks is split into $ S $ message fragments containing $ W_1,\dots,W_S $ data bits, respectively, with $ \sum_{s=1}^{S} W_s = W $. We define $ \{W_1,\dots,W_S\} $ as the data profile. These fragments are distributed over $ S $ $ J $-bit chunks in order. We fix $ W_1 = J $ whereas $ W_s < J $ for $ s=2,\dots,S $. To fill the $ s $-th chunk, the tree encoder appends $ V_s = J - W_s $ parity check bits behind the data bits. These check bits are formed based on the pseudo-random linear combinations of the data bits of the previous blocks. Then at the inner encoding step, each $ J $-bit chunk with index value $ j $ is mapped to the corresponding codeword drawn from the $ j $-th column of the inner codebook.

To decode and reconstruct the message list at the BS, at first, the inner (ML) decoder performs AD to the received signals and obtains a list of emitted $ J $-bit chunks at each time slot, represented by $ \{\mathcal{L}_1,\dots,\mathcal{L}_S\} $. Obviously, most of the message sequences in $ \mathcal{L}_1 \times \mathcal{L}_2 \times \dots \times \mathcal{L}_S $ are unqualified. At the outer tree decoding step, the goal of the tree decoder is to recognize all possible message sequences fitting perfectly the random parity check rules and acquire a transmitted message list $ \mathcal{L} $. The decoder treats the chunks in $ \mathcal{L}_1 $ as the stage-one roots. There are $ \left| \mathcal{L}_2 \right| $ possible choices viewed as leaves to each root at the ensuing stage, but only a few of them satisfy the parity check constraints while others are eliminated, so are the circumstances in the following stages where remaining leaves at the preceding stage will be treated as roots. Eventually, the tree decoder acquires the valid message sequences by traversing all the remaining valid paths from stage $ 1 $ to $ S $.

The performance of the system with concatenated coding scheme is described in terms of per-user probability of misdetection \eqref{pmd} and probability of false-alarm \eqref{pfa} as well.

\section{Simulation Results}

In order to compare with the previous literature, we follow the same system settings as \cite{110}. The massage of length $ W=96 $ is divided into $ S=32 $ blocks of size $ J=12 $ based on the data profile $ \left\lbrace 12,3,3,\dots,3,0,0,0 \right\rbrace $ and encoded by the tree encoder. Then the codewords are separately emitted within $ S $ time slots over $ S $ fading blocks of $ D=100 $ dimensions each.

During the simulation of the iterative algorithm proposed in Section III, we take $ Q=4\times10^4 $ and $ Q_{\text{mod}}=2^J/2 = 2048 $. We define the estimation error of the algorithm as
\begin{equation}
	e_{\bm{\gamma}} = \left\| \bm{\gamma}^q-\bm{\gamma} \right\|_2.
\end{equation}
We present the outcomes in Figure 1, where the red solid line denotes the performance of our CD algorithm with BLA and the blue solid line denotes the CD algorithm with random coordinate selection in \cite{111}. Note that the proposed algorithm reveals better estimation accuracy than the conventional method at the very beginning since the algorithm is most likely to update the coordinates that yield deep descents of $ f(\bm{\gamma}) $ in the first few steps, and for the same reason the estimation error rapidly decreases somewhere along the red curve. These two algorithms achieve the same performance after adequate iterations. It can be observed that the traditional approach converges to the outcome with the minimum error after about $ 3\times10^4 $ iterations while our method takes only approximately $ 1.2\times10^4 $ iterations, indicating that the suggested algorithm leads to a faster convergence rate without losing accuracy.

The average SNR of a generic active user $ k \in \mathcal{K}_{\text{a}} $ over a coherent block is given by
\begin{align}
\varepsilon_k &= \dfrac{1}{N_k}\sum_{n=1}^{N_k}\dfrac{\left\| \mathbf{a}_{i_k} \right\|^2_2g_{k}\mathbb{E}\{\left\| \mathbf{h}_{k,n} \right\|^2_2\}}{\mathbb{E}\{\left\| \mathbf{Z} \right\|^2_\text{F}\}} = \dfrac{g_k}{\sigma^2}
\end{align}
where $ \mathbf{h}_{k,n} $ is the $ n $-th column of the $ M \times N_k $ matrix $ \tilde{\mathbf{H}}_k $. For convenience, we fix the LSFCs of all users to a constant, i.e., $ g_{k} \equiv g $ for $ k \in \mathcal{K}_{\text{tot}} $. We evaluate the system performance by the sum of the two types of error probabilities, i.e., $ P_{\text{e}}=p_{\text{md}}+p_{\text{fa}} $. Let there be $ K_{\text{a}} =300 $ single-antenna active users when we alter the number of antennas at the BS, simulation results are illustrated in Figure 2. If we wish to achieve a total error rate $ P_{\text{e}}<0.05 $, when $ M=300 $, the SNR required for i.i.d. channels is approximately $ 0.2 $ dB and takes the value of $ 1.5 $ dB for correlated channels. When $ M=400 $, these values reduce to $ -3.3 $ dB and $ -2.4 $ dB, respectively. It can be observed that with the growth of $ M $, the SNR needed for reliable communications is immensely decreased, especially when $ M>K_{\text{a}} $. It suggests that the system error probability can be decreased as desired simply by increasing the number of receiving antennas.

\begin{figure}[htb]
	\centerline{\includegraphics[width=10cm]{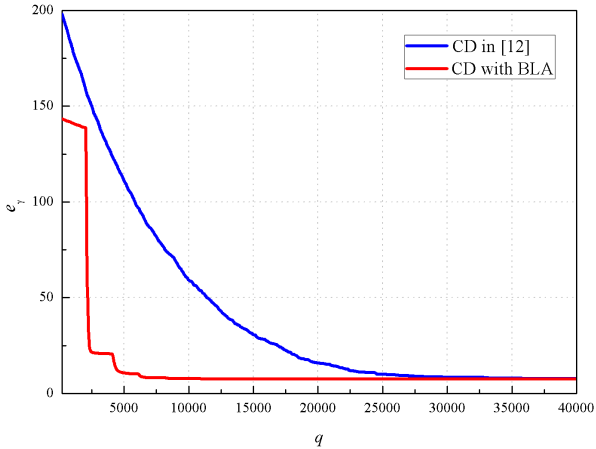}}
	\caption{Estimation errors of two coordinate descent algorithms.}
	\label{fig1}
\end{figure}

\begin{figure}[htb]
	\centerline{\includegraphics[width=10cm]{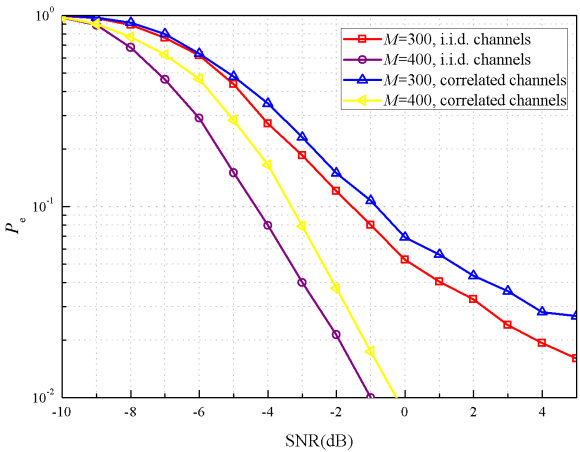}}
	\caption{Error probability as a function of SNR with different values of $ M $. $ K_{\text{a}}=300 $, $ D = 100 $, $ W = 96 $, $ S = 32 $, $ J = 12 $.}
	\label{fig2}
\end{figure}

\section{Conclusion}

In this paper, we investigate the problem of massive connectivity under the scheme of U-RA. In view of realistic propagation environments, we consider a general joint-correlated channel model with LOS components. Facing the problem of signal recovery at the massive MIMO BS, we follow the rationale of the covariance-based recovery problem and introduce the CD algorithm with BLA to conduct AD to the active codewords corresponding to transmitted messages. We point out that the proposed algorithm has a remarkable error to iteration boundary and leads to a faster convergence rate than conventional descent approaches. We take advantage of a coupled coding strategy to further reduce the system complexity. Our simulations indicate that if sufficient numbers of receiving antennas are equipped at the BS, the SNR required for a certain error probability $ P_{\text{e}}=0.05 $ in our scheme suffers from a $ 0.9 $ to $ 1.3 $ dB loss to the results in i.i.d. channels.

\bibliographystyle{IEEEtran}
\bibliography{IEEEabrv,IEEEexample}

\begin{thebibliography}{10}
\providecommand{\url}[1]{#1}
\csname url@samestyle\endcsname
\providecommand{\newblock}{\relax}
\providecommand{\bibinfo}[2]{#2}
\providecommand{\BIBentrySTDinterwordspacing}{\spaceskip=0pt\relax}
\providecommand{\BIBentryALTinterwordstretchfactor}{4}
\providecommand{\BIBentryALTinterwordspacing}{\spaceskip=\fontdimen2\font plus
\BIBentryALTinterwordstretchfactor\fontdimen3\font minus
  \fontdimen4\font\relax}
\providecommand{\BIBforeignlanguage}[2]{{%
\expandafter\ifx\csname l@#1\endcsname\relax
\typeout{** WARNING: IEEEtran.bst: No hyphenation pattern has been}%
\typeout{** loaded for the language `#1'. Using the pattern for}%
\typeout{** the default language instead.}%
\else
\language=\csname l@#1\endcsname
\fi
#2}}
\providecommand{\BIBdecl}{\relax}
\BIBdecl

\bibitem{101}
M.~Ke, Z.~Gao, Y.~Wu, and X.~Meng, ``Compressive massive random access for
  massive machine-type communications ({mMTC}),'' in \emph{Proc. {IEEE} Global
  Conf. Signal. Inf. Process.}, Anaheim, USA, Nov. 2018, pp. 156--160.

\bibitem{102}
L.~Liu and W.~Yu, ``Massive connectivity with massive {MIMO}-{Part} {I}: Device
  activity detection and channel estimation,'' \emph{{IEEE} Trans. Signal
  Process.}, vol.~66, no.~11, pp. 2933--2946, June, 2018.

\bibitem{1021}
L.~Liu, E.~G. Larsson, W.~Yu, P.~Popovski, C.~Stefanovic, and E.~de~Carvalho,
  ``Sparse signal processing for grant-free massive connectivity: A future
  paradigm for random access protocols in the {Internet} of {Thing}s,''
  \emph{{IEEE} Signal Process. Mag.}, vol.~35, no.~5, pp. 88--99, Sept. 2018.

\bibitem{103}
M.~Ke, Z.~Gao, Y.~Wu, X.~Gao, and R.~Schober, ``Compressive sensing based
  adaptive active user detection and channel estimation: Massive access meets
  massive {MIMO},'' \emph{{IEEE} Trans. Signal Process.}, vol.~68, pp.
  764--779, 2020.

\bibitem{104}
Y.~Polyanskiy, ``A perspective on massive random-access,'' in \emph{2017 {IEEE}
  Int. Symp. Inf. Theor. (ISIT)}, Aachen, 2017, pp. 2523--2527.

\bibitem{105}
Y.~Wu, X.~Gao, S.~Zhou, W.~Yang, Y.~Polyanskiy, and G.~Caire, ``Massive access
  for future wireless communication systems,'' \emph{{IEEE} Wireless Commun.},
  vol.~27, no.~4, pp. 148--156, Aug. 2020.

\bibitem{106}
O.~Ordentlich and Y.~Polyanskiy, ``Low complexity schemes for the random access
  gaussian channel,'' in \emph{2017 {IEEE} Int. Symp. Inf. Theor. (ISIT)},
  Aachen, 2017, pp. 2528--2532.

\bibitem{107}
A.~Vem, K.~R. Narayanan, J.~Cheng, and J.~Chamberland, ``A user-independent
  serial interference cancellation based coding scheme for the unsourced random
  access gaussian channel,'' in \emph{2017 {IEEE} Inf. Theor. Workshop (ITW)},
  Kaohsiung, 2017, pp. 121--125.

\bibitem{108}
V.~K. Amalladinne, A.~Vem, D.~K. Soma, K.~R. Narayanan, and J.~Chamberland, ``A
  coupled compressive sensing scheme for unsourced multiple access,'' in
  \emph{2018 {IEEE} Int. Conf. Acoust. Speech Signal Process. (ICASSP)},
  Calgary, AB, 2018, pp. 6628--6632.

\bibitem{109}
A.~Fengler, P.~Jung, and G.~Caire, ``{SPARCs} and {AMP} for unsourced random
  access,'' in \emph{2019 {IEEE} Int. Symp. Inf. Theor. (ISIT)}, Paris, France,
  2019, pp. 2843--2847.

\bibitem{110}
A.~Fengler, S.~Haghighatshoar, P.~Jung, and G.~Caire, ``Grant-free massive
  random access with a massive {MIMO} receiver,'' in \emph{Conf. Rec. Asilomar
  Conf. Signals Syst. Comput. (ACSSC)}, Pacific Grove, CA, USA, 2019, pp.
  23--30.

\bibitem{111}
S.~Haghighatshoar, P.~Jung, and G.~Caire, ``Improved scaling law for activity
  detection in massive {MIMO} systems,'' in \emph{2018 {IEEE} Int. Symp. Inf.
  Theor. (ISIT)}, Vail, CO, 2018, pp. 381--385.

\bibitem{204}
D.-S. Shiu, G.~J. Foschini, M.~J. Gans, and J.~M. Kahn, ``Fading correlation
  and its effect on the capacity of multielement antenna systems,''
  \emph{{IEEE} Trans. Commun.}, vol.~48, no.~3, pp. 502--513, Mar. 2000.

\bibitem{201}
X.~Gao, B.~Jiang, X.~Li, A.~B. Gershman, and M.~R. McKay, ``Statistical
  eigenmode transmission over jointly correlated mimo channels,'' \emph{{IEEE}
  Trans. Inf. Theory}, vol.~55, no.~8, pp. 3735--3750, Aug. 2009.

\bibitem{202}
V.~Raghavan and A.~M. Sayeed, ``Sublinear capacity scaling laws for sparse
  {MIMO} channels,'' \emph{{IEEE} Trans. Inf. Theory}, vol.~57, no.~1, pp.
  345--364, Jan. 2011.

\bibitem{203}
S.~Ghosh, B.~D. Rao, and J.~R. Zeidler, ``Techniques for {MIMO} channel
  covariance matrix quantization,'' \emph{{IEEE} Trans. Signal Process.},
  vol.~60, no.~6, pp. 3340--3345, June 2012.

\bibitem{205}
J.-P. Kermoal, L.~Schumacher, K.~I. Pedersen, P.~E. Mogensen, and
  F.~Frederiksen, ``A stochastic {MIMO} radio channel model with experimental
  validation,'' \emph{{IEEE} J. Sel. Areas in Commun.}, vol.~20, no.~6, pp.
  1211--1226, Aug. 2002.

\bibitem{206}
A.~M. Sayeed, ``Deconstructing multiantenna fading channels,'' \emph{{IEEE}
  Trans. Signal Process.}, vol.~50, no.~10, pp. 2563--2579, Oct. 2002.

\bibitem{207}
W.~Weichselberger, M.~Herdin, H.~Ozcelik, and E.~Bonek, ``A stochastic {MIMO}
  channel model with joint correlation of both link ends,'' \emph{{IEEE} Trans.
  Wireless Commun.}, vol.~5, no.~1, pp. 90--100, Jan. 2006.

\bibitem{301}
S.~Bubeck and N.~Cesa-Bianchi, ``Regret analysis of stochastic and
  nonstochastic multi-armed bandit problems,'' \emph{Found. Trends Mach.
  Learn.}, vol.~5, pp. 1--122, Dec. 2012.

\bibitem{302}
O.-C. Granmo, ``Solving two‐armed {Bernoulli} bandit problems using a
  {Bayesian} learning automaton,'' \emph{Int. J. Intell. Comp. Cybern.},
  vol.~3, no.~2, pp. 207--234, 2010.

\end{thebibliography}

\end{document}